\documentclass[twocolumn,amsmath,amssymb,prl,footinbib,floatfix,superscriptaddress]{revtex4}

\usepackage{graphicx}
\usepackage{dcolumn}
\usepackage{bm}

\usepackage{helvet}
\usepackage{amssymb}
\usepackage{amsmath}
\usepackage{amsfonts}
\usepackage{amssymb}
\usepackage{hyperref}
\usepackage{color}
\usepackage{soul}

\usepackage{xcolor}
\definecolor{red}{rgb}{0.7,0,0}
\definecolor{green}{rgb}{0.,0.35,0.}
\definecolor{blue}{rgb}{0.2,0.2,0.7} 
\definecolor{black}{rgb}{0.15,0.15,.15}

\def\barray{\begin{eqnarray}}
\def\earray{\end{eqnarray}}
\def\beq{\begin{equation}}
\def\eeq{\end{equation}}

\makeindex

\begin{document}

\title{Equipartition of the Entanglement Entropy}

\date{\today}


\author{J.~C.~Xavier }

\affiliation{Instituto de F\'{\i}sica, Universidade Federal de Uberl\^andia,
  C. P. 593, 38400-902 Uberl\^andia, MG, Brazil}

\author{F. C. Alcaraz}

\affiliation{Instituto de F\'{\i}sica de S\~ao Carlos, Universidade de S\~ao
  Paulo, Caixa Postal 369, S\~ao Carlos, SP, Brazil}

\author{G. Sierra}
\affiliation{Instituto de F\'{\i}sica Te\'orica  UAM/CSIC, Universidad Aut\'onoma de Madrid,  Cantoblanco 28049 , Madrid, Spain}
\affiliation{Kavli Institute for Theoretical Physics, University of California, Santa Barbara, CA 93106, USA}

\begin{abstract}

The entanglement in a  quantum system that possess  an internal  symmetry,  characterized
by the  $S^z$-magnetization  or $U(1)$-charge,  is distributed among different sectors. 
 The aim  of this letter is to gain a deeper understanding of the contribution to the 
 entanglement entropy in each of those sectors for the ground state of conformal
 invariant critical  one dimensional systems.
We find surprisingly  that the entanglement entropy
is equally distributed  among the  different magnetization  sectors. 
Its value is given by the standard area law violating logarithmic term, that depends on the central
charge $c$, minus  a double logarithmic 
correction related to  the zero temperature  susceptibility.  This result provides  a  new method   to estimate
simultaneously the central charge $c$ and the critical  exponents of
$U(1)$-symmetric quantum chains.   
The method  is numerically simple and    
gives precise  results for the spin-$\frac{1}{2}$ quantum  XXZ chain.
We also compute the probability distribution 
of the magnetization in  contiguous  sublattices. 
\end{abstract}













\maketitle
 
 
{\it Introduction}. In recent years  the study of entanglement in quantum many body systems,  and in quantum field theory,  
has been carried out  intensively. As a result of it, many links have  been established among previously disconnected
areas of Physics, Computer Science and Mathematics. These studies  have led to  a quantum
information perspective of phase transitions and topological order, topics that belonged   traditionally 
to Condensed Matter Physics and Statistical Mechanics \cite{ent}. 
For most of the quantum  critical  systems in one spatial dimension, a precise characterization of entanglement has 
been achieved  thanks to the powerful methods of  Conformal Field Theory (CFT).  In these systems, 
the area law of the von Neumann entanglement entropy (EE) of the ground state (GS), in a single interval \cite{area}, 
develops  a logarithmic violation parameterized 
by the central charge $c$ of the underlying CFT \cite{CW94}-\cite{K04}.

Employing ultra-cold atoms loaded in  optical lattices,  it is nowadays  possible to simulate  many 
one-dimensional quantum systems \cite{review-OL}. Quite  recently,  a measurement of entanglement was done 
using  a one-dimensional optical lattice composed of a  few $^{87}$Rb atoms
\cite{exp-EE}. Since the number of  atoms involved  in these experiments is small, finite-size effects play an important
role in measuring the EE. Fortunately,  CFT predicts  the leading finite-size
correction of the R\'enyi entropy of the GS of a chain of $L$ sites, which is given by 
  \cite{CW94}-\cite{K04}
\begin{equation}\label{CC}
 S^{(n)}_{A,CFT} =c_n^{(b)}+\frac{c}{3 b}\left(1+\frac{1}{n}\right)\ln\left[\frac{b L}{\pi}\sin\frac{\pi  x}{L}\right],
\end{equation}
where $x$ is the size of the subsystem $A$,
 $b=1,2$ for periodic/free boundary conditions (PBC/FBC);
and $c_n^{(b)}$ is a non universal constant. 
The EE corresponds to the choice  $n=1$.


Besides the central charge $c$,  the entanglement properties 
of a quantum chain can also depend on  the critical exponents or the operator 
content of the underlying CFT. This dependence was previously  observed   in 
 the entanglement for  multiple intervals \cite{2int-refs},
scaling  corrections \cite{corrections}, parity effects \cite{parity},  and 
in  the primary states and descendants of the CFT \cite{exc-stat}.

The aim of this Letter is to split the total entanglement 
into the contributions coming from disjoint symmetry  sectors. 
We carry out this  analysis for the  
critical quantum chains that have a  $U(1)$ symmetry that,  in the scaling limit, 
develops  a $U(1)$ Kac-Moody algebra  (KM).
Those are the models with a critical line with 
 continuously varying exponents \cite{KB}. 
As a byproduct of our calculation, we present herewith 
a simple new  method to evaluate simultaneously the central charge and  the critical exponents 
for this class of  quantum chains. We hope, that these results
could be tested in ultracold atoms experiments as the ones
performed in reference  \cite{exp-EE}.

Let us start with a general quantum chain with $L$ sites, whose  Hamiltonian  $\hat{H}= \sum_{i=1}^L h_{i,i+1}$,
commutes with the  magnetization  operator $S^z = \sum_{i=1}^L S^z_i$,
 where $S^z_i$ are spin $s-$matrices.
  Let $| \psi \rangle$ be a common  eigenstate of $\hat{H}$ and $S^z$, with eigenvalues 
 $E$ and $M$ respectively. We split the chain into  disjoints  blocks  $A \;(i=1, \dots, x)$ and   $B \; (i=x+1, \dots, L)$, 
 and compute the reduced density matrix $\rho_A = {\rm tr}_{B} \;  \rho \;( \rho=   | \psi \rangle \langle \psi|$). 
 The magnetization operator also splits into the sum $S^z = S^z_A + S^z_B$.  Then, tracing over the Hilbert
 subspace of  the block $B$  in the equation $[S^z,  \rho ]= 0$
yields  $[S^z_A, \rho_A] =0$.  This implies 
\beq
\rho_A =\oplus_m   \, \tilde{\rho}_{A,m}= \oplus_m \, p_{A,m} \, \rho_{A,m} \,  , 
\label{1}
\eeq 
where $- sx \leq m \leq sx$,  ${\rho}_{A,m}$ is a density matrix with eigenvalue $m$ of $S^z_A$, and 
$p_{A,m}={\rm tr}\tilde{\rho}_{A,m} \geq 0$ is the probability of finding 
$m$ in a measurement of $S^z_A$.

The decomposition (\ref{1})   is implemented  normally  in  numerical methods, like the DMRG \cite{W92} 
and MPS, MERA, etc \cite{MPS}  to reduce the memory resources needed for  high precision results. 
The latter equation  implies  
\beq 
S_A = \sum_m p_{A,m} \, S_{A,m} + H_A   \, , 
\label{2}
\eeq 
where $S_A= - {\rm tr} \rho_A \ln \rho_A$,  $S_{A,m}= - {\rm tr} \rho_{A,m} \ln \rho_{A,m}$,
and $H_A = - \sum_{m} p_{A,m} \ln p_{A,m}$. 
Equation (\ref{2}) means that the quantum entropy 
in the  subsystem $A$ is greater, in general,  
to the weighted sum of the entropies of the different magnetization sectors. 
This fact expresses  the  holistic nature of quantum  entanglement. 
Actually,  Eq. (\ref{2})  can be seen as a special case of the general Holevo theorem in
quantum information theory \cite{holevo,bookNielsen}, that states that the maximum
information we can extract from a general mixed state,
$\rho=\sum_m p_m \rho_m$, is given by the difference
$S(\rho)-\sum_mp_mS(\rho_m)$. In the case of Eq. (\ref{2})  the maximum information is given by the Shannon entropy.

The critical chains studied in this Letter, reveals surprisingly that the contributions 
$S_{A,m}$ to the entropy $S_A$, are equal for  the low  values of the magnetization $|m| $. 
We call  this situation {\em equipartition of entanglement entropy}.   


{\it Analytic predictions}. 
The  previous discussion applies to any quantum lattice  system with a $U(1)$ symmetry. 
In the following we shall derive  analytic predictions  of $p_{A,m}$  and 
$S_{A,m}$, for  critical spin-$s$ quantum  chains, 
like the spin-$s$ XXZ model. 
In  a block $A$ with $x$ sites, 
Eq. (\ref{1}) can be inverted to  obtain 
\beq
p_{A,m} \rho_{A,m} = \frac{1}{ 2 s x +1} \sum_{n = - s x}^{s x} e^{ \frac{2 \pi i n}{ 2 s x+1} (S^z_A  - m) } \, \rho_A \, , 
\label{3}
\eeq  
where  the sum projects the density matrix  $\rho_A$ into the sector with $S^z_A  =m$.
In the limit $x \gg 1$, Eq. (\ref{3}) becomes 
\beq
p_{A,m} \rho_{A,m} =  \int_{-1/2}^{ 1/2} d \phi  \;   e^{ 2 \pi i  \phi  (S^z_A - m) } \, \rho_A \, . 
\label{4}
\eeq  
Taking the trace over the states in $A$, and using ${\rm tr}_A   \rho_A  = {\rm tr}_A   \rho_{A,m} =1$,  
gives the probability distribution 
\beq
p_{A,m}  =  \int_{-\frac{1}{2}}^{ \frac{1}{2} } d \phi  \;   e^{ - 2 \pi i  m \phi }    {\rm tr}_A  ( e^{ 2 \pi i  \phi S^z_A   } \,  \rho_A )  \, . 
\label{5}
\eeq  
Similarly, the $n^{\rm th}$ power of Eq. (\ref{4}) yields
\barray 
\label{6}
p^n_{A,m} {\rm tr}_A  \; \rho_{A,m} ^n  & = &  \prod_{j=1}^n  \int_{- 1/2}^{1/2} d \phi_j  \; e^{ - 2 \pi i  m     \sum_{j=1}^n  \phi_j}    \\
& & 
  \times \;   {\rm tr}_A  (  e^{  2 \pi i   S^z_A      \sum_{j=1}^n  \phi_j}   \rho^n_A  )  \, ,  \nonumber 
\earray 
that together with (\ref{5}),  provides the R\'enyi entropies 
$S_{A,m}^{(n)} = \frac{1}{1-n} \log  {\rm tr}_A \rho^n_{A,m}$. 
To  find  ${\rm tr}_A  (  e^{  2 \pi i  \phi    S^z_A    }   \rho^n_A  )$, 
we  extend   the general  formalism to construct  the  entanglement Hamiltonian
in CFTs  \cite{CT16}, that we summarize 
below. 

The reduced density matrix $\rho_A$ in CFT is given by
\beq
\rho_A =  \frac{1}{Z_1} e^{- 2 \pi K_A}, \quad   Z_1 = {\rm tr}_A \;  e^{ - 2 \pi K_A}  \, , 
\label{7}
\eeq 
where $K_A = \int_A dx \, T_{00}(x)/f'(x)$ is the entanglement Hamiltonian and  $T_{00}$ is
a component of the stress tensor \cite{CT16}.  $f(z)$  is the conformal map from the euclidean space-time, 
with a cut along the interval $A$ and two boundaries, into  an annulus of width $W$ and height $2 \pi$.   
Taking the trace of the $n^{\rm th}$ power in (\ref{7}) yields
\beq
{\rm tr}_A \,  \rho_A^n = \frac{Z_n}{Z_1^n} , \quad   Z_n = {\rm tr}_A \;  e^{ - 2 \pi n  K_A}   \, , 
\label{8}
\eeq 
where $Z_n$ is the euclidean  partition function of an $n-$sheeted cover of the original 
space-time with conical singularities around the end points of $A$.  
We propose the following extension  of  Eq. (\ref{8})  to  CFTs  with  a $U(1)$ KM symmetry: 
\beq
{\rm tr}_A \, \left( e^{ 2 \pi i \phi J_0} \,   \rho_A^n \right)  =   \frac{Z_n(\phi)}{Z_1^n} =
\frac{ {\rm tr}_A  \left( e^{ 2 \pi i \phi J_0}   e^{ - 2 \pi n  K_A} \right)}{ Z_1^n}  \, ,
\nonumber 
\eeq 
where $J_0$ is the zero mode of the $U(1)$ current $J(z)$. 
$Z_n(\phi)$ is the  partition function, given  in Eq. (\ref{8}),  but  with fugacity $2 \pi i \phi$.
 Since the  eigenvalues of $K_A$ are given by $\pi( \Delta_{p,m} - c/24)/W$,
where $\Delta_{p,m}$ are the dimensions of the boundary operators \cite{CT16} and $m$ is the eigenvalue of $J_0$, 
we obtain 
\beq
Z_n(\phi)  = q^{ - n c/24} \sum_{p,m} d_{p,m}  \, q^{ n \Delta_{p,m}}  e^{ 2 \pi i m \phi}  \, ,
\label{10}
\eeq 
where $q = e^{ - 2 \pi^2/W}$ and $d_{p,m}$ is the degeneracy of the boundary operator $(p,m)$. 
In the case of the ground state of the CFT, with  periodic/free  boundary
conditions, the width $W$ should be fixed to [recall Eq. (\ref{CC})]:  
\beq
W = \frac{2}{b}   \ln (b L_c(x)),  
\quad L_c(x) = \frac{ L}{\pi} \sin \frac{ \pi x}{L} \,.
\label{11}
\eeq
For the thermal state at temperature  $1/\beta$, 
$W = 2 \ln (\frac{ \beta}{ \pi} \sinh \frac{\pi x}{\beta} )$ \cite{CT16}. 

The first application  of the analytic formula  (\ref{10}) is the Luttinger liquid 
which is a CFT with  $c=1$ and a $U(1)$ symmetry generated by 
the current operator $J(z) = i \sqrt{K} \partial \varphi(z)$,
where $\varphi(z)$ is a chiral boson and $K$ a constant. 
A state with charge $m \in  Z + a$ (with $a=0, \frac{1}{2})$  is associated to the vertex operator
$e^{ i m \varphi(z)/\sqrt{K}}$, and has conformal weight
$h_m =   m^2/(2 K)$.   
The partition function (\ref{10}) reads in this case 
\beq
Z_n( \phi) = \frac{1}{ \eta(q^n)} \sum_{m \in Z + a} q^{  \frac{n m^2}{2K} } e^{ 2 \pi i m \phi} 
=  \frac{\theta_{a,0} ( \phi , \frac{ n \tau}{K})}{ \eta(q^n)} 
  \, , 
\label{12}
\eeq 
where $\tau = i\pi/ W$,  
$\eta(q) = q^{\frac{1}{24}} \prod_{n=1}^\infty ( 1 - q^n)$ is the Dedekind eta function, 
and  $\theta_{a,c}(z,\tau) = \sum_{n \in Z} e^{ \pi i \tau (n+a)^2 + 2 \pi i n (z+c)}$ is 
a Jacobi theta function with characteristics. 
In the limit $L \gg 1$, one has
$W \gg  1$ and therefore  $q \sim 1$, so that a large number of terms 
contribute to  Eq. (\ref{12}).  However, 
using the modular transformation $\tau \rightarrow - 1/\tau$ \cite{NS}, 
\beq
\theta_{a,0}(z, \tau) = \sqrt{ \frac{i}{ 2 \tau}}  e^{ - i  \frac{z^2}{2 \tau}}   \sum_{a'=0, \frac{1}{2}}  e^{ 4 \pi i a a'}  \theta_{a',0} (- \frac{z}{\tau}, - \frac{2}{ \tau})
\label{x}
\eeq 
and $\eta(-1/\tau) = \sqrt{ \tau/i} \;  \eta(\tau)$,  we  obtain 
\beq
Z_n(\phi) \sim e^{ \frac{W}{n} ( \frac{1}{12} - K  \phi^2) }  \, . 
\label{13}
\eeq  
For special values of  $K$, the CFT is rational and  $Z_n(\phi)$ 
becomes  a finite sum  (e.g.  if $K^2$ is an even number \cite{yellow}) 
\beq
Z_n(\phi)  =  \sum_{j}  \, n^j  \,   \chi_j(q^n, \phi)   \, ,
\label{14}
\eeq 
where $n^j$ are non-negative integers, that depend  on 
 the boundary conditions on the annulus.   The coefficients 
$\chi_j(q^n, \phi)$ are denoted non-specialized characters
that are labelled by  the representation $j$ of an extended 
KM algebra.   Their  modular
transformations  \cite{non-spe,yellow}, 
have  been used to study the correlators in  the  multichannel Kondo model \cite{AL94}
and bulk susceptibilities  \cite{sus,B83}.   

The second application we report  deals with the spin $s$-isotropic 
exactly solvable model \cite{hspin}. This is a critical  system    
described by the Wess-Zumino-Witten (WZW) model  $SU(2)_k$ at   level  $k = 2 s$, and central charge
 $c= \frac{ 3 k}{k+2}$. 
The  model  contains a similar 
  $U(1)$ current   operator,  which is now $J^z(z) = i \sqrt{\frac{k}{2}}  \partial \varphi(z)$.
  The primary fields are labelled by the total spin $j=0, \frac{1}{2}, \dots, \frac{k}{2}$.
The partition function (\ref{14}) is a linear
combination of the non-specialized characters $\chi_j(q, \phi)$
of $SU(2)_k$ and using their   modular transformations   \cite{yellow,non-spe}, 
we obtain \cite{future} 
\beq
Z_n(\phi) \sim e^{ \frac{W}{n} ( \frac{c}{12} - \frac{k}{2}   \phi^2) }. 
\label{15}
\eeq  
For the  spin-1/2 chain, $c=1$ and $k=1$, and then (\ref{15}) coincides with 
(\ref{13}), for   $K=\frac{1}{2}$, where 
Luttinger liquid has an enhanced  $SU(2)_1$ symmetry. 

We can  summarize both previous  applications as 
\beq
{\rm tr}_A \, \left( e^{ 2 \pi i \phi J_0} \,   \rho_A^n \right)  \propto 
e^{ W [  \frac{c}{12}  \left( \frac{1}{n} - n \right)  - \frac{{\cal  K}}{n}  \phi^2]  }  \, , 
\label{18}
\eeq 
where  ${\cal K} = K$ or ${\cal K} = \frac{k}{2}$ for the Luttinger liquid
and the spin-$s$ chain, respectively. 
Quite remarkably, this parameter  satisfies 
the {\em universal relation}  ${\cal K} =  \pi v_s \chi$,
where  $v_s$ is the   spin wave velocity, and $\chi$ is   the 
zero field susceptibility at zero temperature  of these distinct   spin chains \cite{sus,B83}.
The case $n=1$ in Eq. (\ref{18}) coincides with  the full
counting statistics (FCS)  for the  subsystem  magnetization $S^z_A$ \cite{FCS,ent-fluc}.
Taking $n>1$ provides a generalized FCS where the entanglement properties
are taken into account.

Let us derive some  consequences of Eq. (\ref{18}) 
for  finite chains with PBC [i.e. $b=1$ in (\ref{11})].  
For  $n=1$, we obtain   the probability distribution 
\beq
p_{A,m}  =  \int_{-\frac{1}{2}}^{ \frac{1}{2} } d \phi  \;   e^{ - 2 \pi i \phi m - \kappa  \phi^2}, \; 
\kappa = 2 {\cal K}  \ln[ g L_c(x)]  \, . 
\label{19}
\eeq 
The constant $g$ comes from the lattice cutoff in the chains, 
that has not been included  in (\ref{18}).  
The highest probability corresponds to $m=0$, 
\beq
p_{A,0}  =  \sqrt{ \frac{ \pi}{ \kappa}}   \; {\rm erf} \left( \sqrt{ \kappa} /2 \right), 
\label{20}
\eeq 
where ${\rm erf}(x)$ is the error function. 
$p_{A,m}$ can be approximated  by
replacing  the integration  limits  in (\ref{19})  by $\pm \infty$, 
\beq
p_{A,m}  \simeq   \sqrt{  \frac{ \pi}{ \kappa} } e^{ - ( \pi m)^2/\kappa}   \, , 
\label{21}
\eeq 
which is a distribution whose   Shannon entropy 
\beq
H_A \sim   \frac{1}{2}  \ln \left( 2  {\cal K}   \ln [g  L_c(x) ] \right)  \, , 
\label{22}
\eeq 
quantifies our knowledge after measurement of sublattice magnetization. 
For a half block it will go as $\ln \ln L$,  a remarkably slow increase with
$L$.
It is interesting to observe that the relation 
${\cal K} = \pi  v_s \chi$ obtained in Refs. \onlinecite{sus} and \onlinecite{B83} , can be derived from
  Eq. (\ref{21}). The zero field  susceptibility  is given by  $\chi=  \langle
  m^2 \rangle/( x T)$, 
where $m$ is the magnetization of a region  of lenght $x$ and $T$ is the
temperature. Using Eq. (\ref{21}), one finds $\langle m^2 \rangle =  \frac{
  \kappa}{2 \pi^2}$, where $\kappa = {\cal K} W=2 \ln (\frac{ \beta}{ \pi}
\sinh \frac{\pi x}{\beta} )$ [Notice that the expression of $\kappa$, defined
  in Eq. (\ref{19}) is for $T=0$, where $W=2  \ln L_c(x)$ ]. In the limit
$x\gg\beta$ one finds  $\langle m^2 \rangle\simeq {\cal K}xT/\pi $, that gives
the relation ${\cal K} = \pi  v_s \chi$.  The same sort of computation provides, for example, the Gibbs entropy for
the subsystem of size $x$, that is given by $S_A=\frac{\pi c}{3}T$.
Note also that since  Eq. (\ref{18}) is related with  the zero field
susceptibility (which  is related with the spin fluctuations), we would
expect a connection between the entanglement and spin fluctuations. This is
very interesting, since measurement of fluctuations are easier to do than the
entanglement ones. Indeed, recently some authors have made this connection
 \cite{ent-fluc} (see also Ref. \onlinecite{suscep-vedral-zoller}).

In the limit $\kappa \gg 1$, the  R\'enyi entropies, 
can be found using (\ref{6}) and (\ref{18}), and behave  asymptotically as 
\beq
\frac{ {\rm tr} \; \rho_{A,m}^n }{ {\rm tr} \; \rho_A^n} \propto \kappa^{ \frac{1}{2} (n-1)}  \, , 
\label{23}
\eeq 
that implies 
\beq
S_{A,m}^{(n)} \simeq   S^{(n)}_{A,CFT} -   \frac{1}{2} \ln \kappa. 
\label{24}
\eeq 
Hence,  the EE of the  density matrix  $\rho_{A,m}$ is dominated by the
EE of  the full density matrix $\rho_A$, with a  reduction 
 $- \frac{1}{2} \ln (2 {\cal K}   \ln[ g L_c(x))]$ that is independent of the quantum number $m$. 
 This is the {\em equipartition} of the EE mentioned above.

{\em Numerical tests.} We have considered  the spin-1/2 XXZ Hamiltonian with PBC  
 \beq
 H=   \sum_{n=1}^L  \left( S^x_{n} S^x_{n+1} + S^y_n  S^y_{n+1} + \Delta S^z_{n} S^z_{n+1} \right)  \, , 
 \label{n1}
 \eeq
 in the critical regime $-1 <   \Delta \leq  1$, whose low energy 
 is described by a Luttinger liquid with parameter $K = \frac{1}{2} ( 1- \frac{1}{\pi} {\rm arcos}(\Delta) )^{-1}$ \cite{,KB,XXZ}. 
 Using the DMRG method we obtained the GS and 
 the reduced density matrices $\rho_A$ and $\rho_{A,m}$.  
We consider system sizes up to $L=600$ under PBC and  
keeping up to $\tilde{m}$=3000 states per block
in the final sweep. We have done $\sim 6-10$ sweeps, and the discarded
weight was typically $10^{-10}-10^{-12}$ at that final sweep.
To verify 
 Eq. (\ref{18}), we  write it as
\beq
\ln  {\rm tr} ( e^{ 2 \pi i \phi S^z_A} \, \rho_A^n )  = - \gamma_n(\phi) \ln( g_n  L_c(x))  + d_n  \, , 
 \label{n2}
 \eeq 
 where $g_n, d_n$  are non universal constants ($g_1 \equiv g$), and 
 \beq 
 \gamma_n(\phi)  =    \alpha_n + \beta_n(\phi),  \; 
 \alpha_n   =   \frac{c}{6} \left( n - \frac{1}{n} \right), \; \beta_n (\phi) = \frac{ 2 {\cal K}}{n} \phi^2 \, . 
 \label{n3} 
 \eeq
In our opinion Eqs. (\ref{n2})-(\ref{n3}) give us the most simple and 
numerically easier method to evaluate the central charge and the Luttinger 
parameter from reduced density matrices. 
The evaluation of  $\rho_{A,m}$
with the DMRG does not require any additional numerical effort because it is already calculated in
the evaluation of $\rho_A$. 
For $n=1$, Eq. (\ref{n2}) yields a  $\phi$-{\it extension} of 
the trace that provides  the Luttinger parameter
 ($\alpha_1=0$, $\beta_1=2 {\cal K}  \phi^2$), and for  $n>1$ it gives 
 the central charge [$\alpha_n = c(n-1/n)/6$].

\begin{figure}[tbp]
\hspace*{-0.5cm}\begin{centering}\includegraphics[width=5.2cm,height=5.2cm]{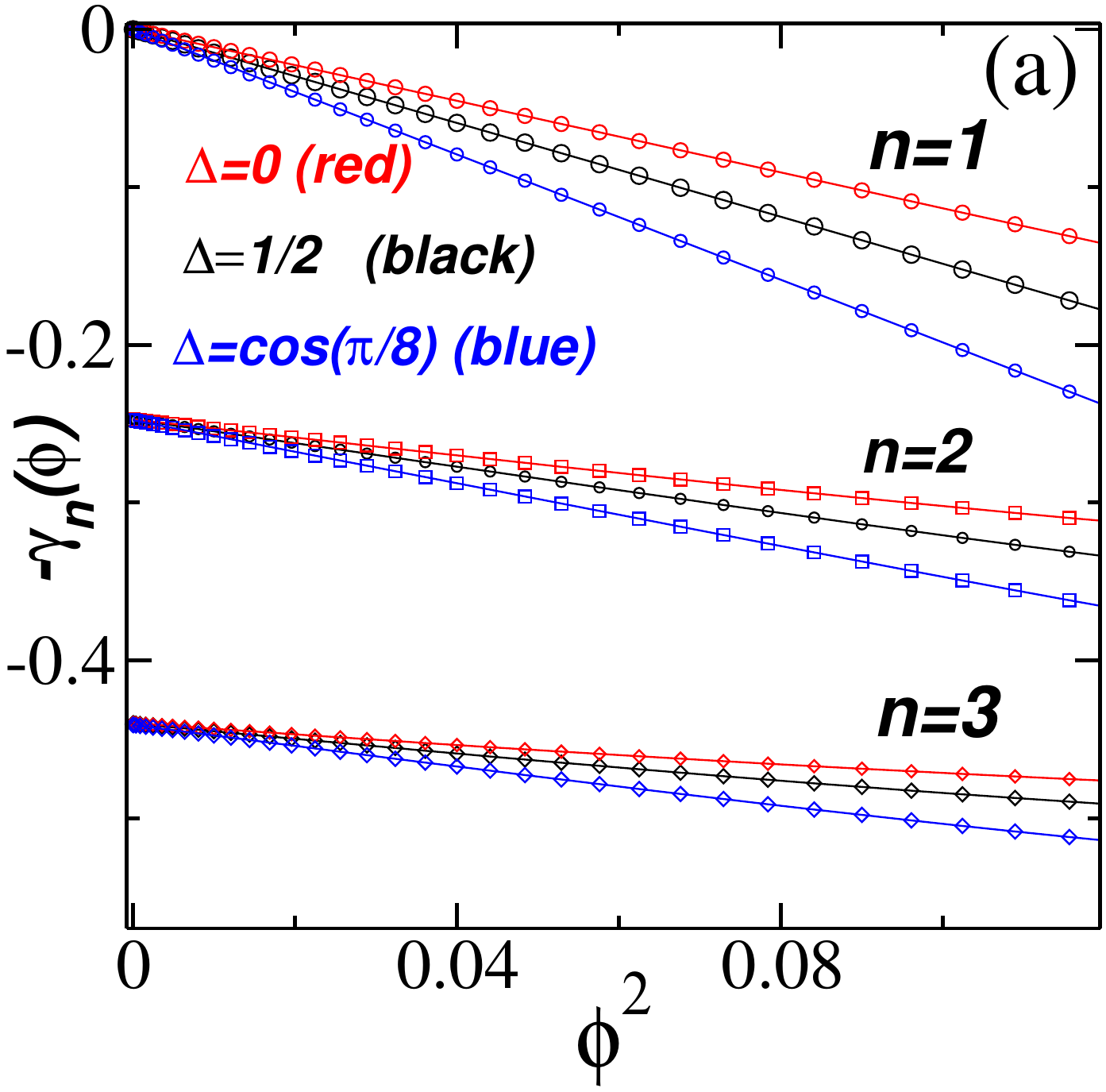}\par\end{centering}
\hspace*{-0.5cm}\begin{centering}\includegraphics[width=5.2cm,height=5.2cm]{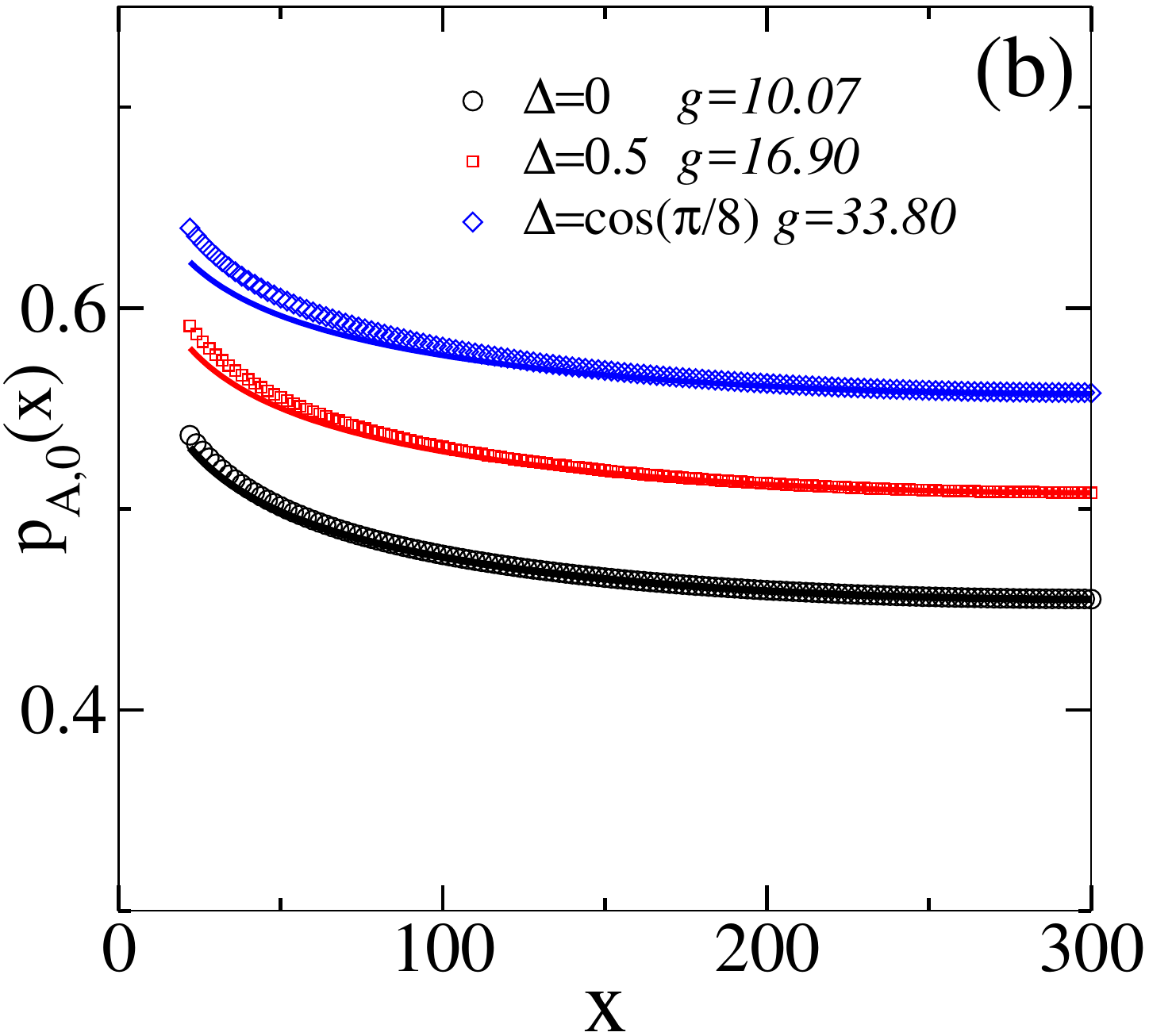}\par\end{centering}
\hspace*{-0.5cm}\begin{centering}\includegraphics[width=5.2cm,height=5.2cm]{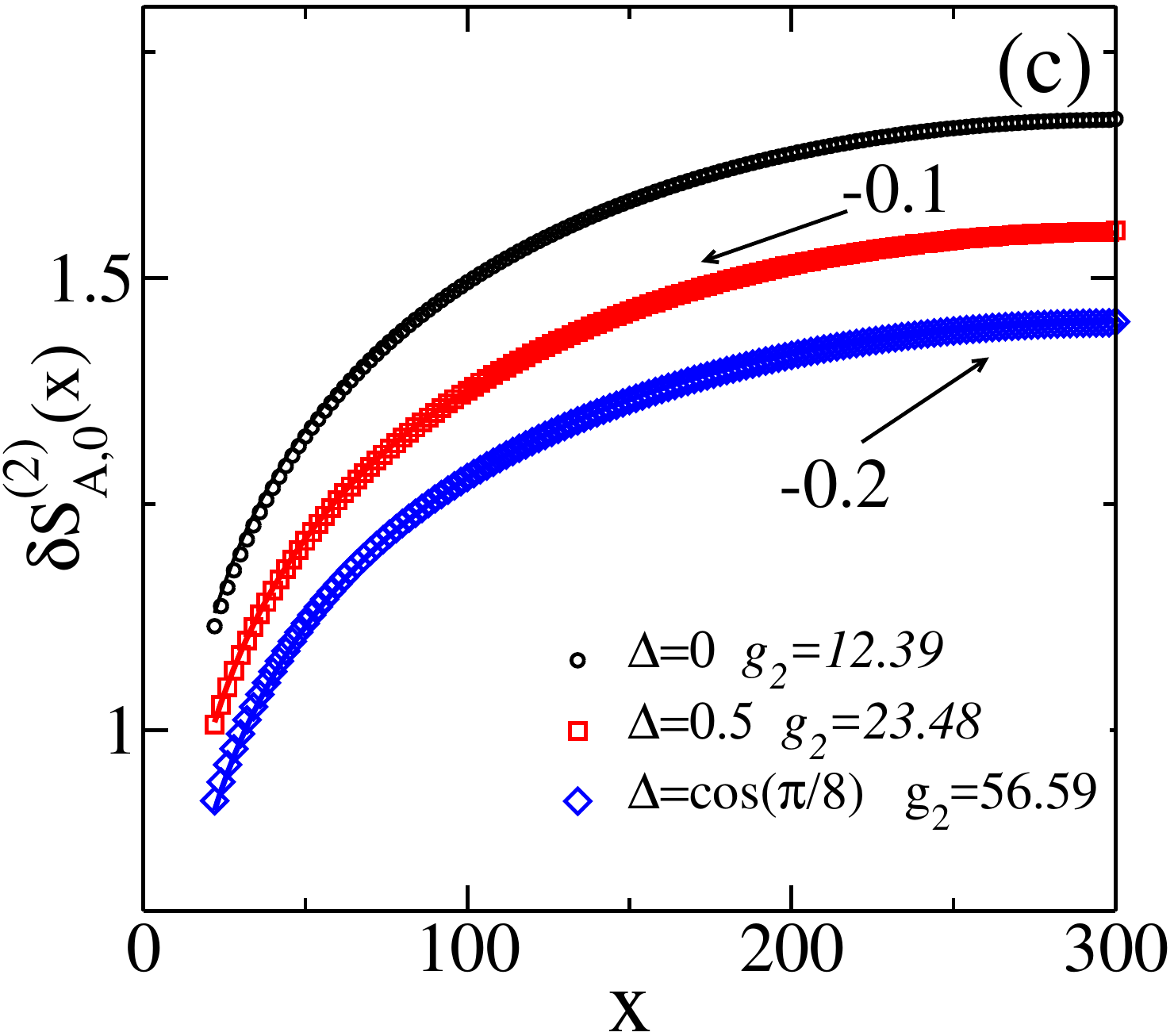}\par\end{centering}
\caption{ (Color online) DMRG results for XXZ quantum chain  for  several
 values of  $\Delta$ and a chain of $L=600$ sites. 
 (a) $\gamma_n(\phi)$ vs. $\phi^2$  for $n=1,2,$ and 3. The data where obtained considering, for each  $\phi$, the sublattice sizes $x \in [80, 300]$.   
 (b)  $p_{A,0}(x)$ vs $x$.  (c) 
  $ \delta S_{A,0}^{(2)} $ vs. x. The numbers $-0.1, -0.2$ are vertical shifts to facilitate the picture. 
  The values of $g$ and $g_2$ in Figs. (b) and (c) are obtained from fitting 
  the data of Fig. (a) to Eq. (\ref{n2}).   
  The symbols are the numerical data and the lines in (b) and (c) are the theoretical
  predictions.
} 
\label{fig1}
\end{figure}

The DMRG data show clearly, for each $\phi$, the linear dependence on  $\ln
L_c(x)$ in Eq. (\ref{n2}). 
We illustrate in Fig. \ref{fig1}(a) the  function $\gamma_n(\phi)$ obtained
from Eq. (\ref{n2}) for several  values of $\Delta$ and $n$. 
 Table \ref{tabI} summarizes  the 
results  for  the estimated  values of  $\alpha_n$  and $\beta_n(\phi)$. Notice the excellent agreement
between the numerical and theoretical results.

  We also test Eq. (\ref{20})  for the XXZ spin-1/2 chain using $g$
  as a fitting parameter. In the case of the XX model, the exact
  value is given by $g = 2 e^{1 + \gamma} = 9.68 \dots$ \cite{future}. 
Fig. \ref{fig1}(b)  illustrates the excellent agreement between
 the numerical and the analytical prediction (\ref{20}) for  $p_{A,0}$ as function of $x$ for three values
of $\Delta$.

 Finally, we present the results for the R\'enyi-2 entropy $S_{A,0}^{(2)}$.
We found that   $S_{A,0}^{(2)}= S_{A}^{(2)}- f(\kappa,\kappa_2)$ \cite{future},
where
 \beq
  f(\kappa,\kappa_2)=   \ln \left[ \frac{  \kappa}{\pi \kappa_2} \frac{ \left[ \sqrt{ 2 \pi \kappa_2} \, 
 {\rm erf} \left( \frac{\sqrt{ \kappa_2}}{2} \right) -2+2e^{- \kappa_2/2}\right] }{\left[ {\rm erf} \left( \frac{\sqrt{ \kappa}}{2}
 \right) \right]^{2}} \right] 
\nonumber 
\eeq 
and 
$\kappa_2 = 2 K \ln [ g_2 L_c(x)]$. The asymptotic behavior  was already  shown in (\ref{24}). 
Fig.  \ref{fig1}(c) shows the DMRG data for  $\delta S_{A,0}^{(2)}=S_{A,0}^{(2)}+f(\kappa,\kappa_2)=c/4\ln[2L_c(x)]+c_2^2$ in the XXZ
spin-1/2 chain with $L=600$,  where we use the values
of $g_1$ and $g_2$ found in Fig. \ref{fig1}(a), shown in the insets of Figs. \ref{fig1}(b,c).

\begin{table}
\begin{tabular}{ccllll}
\hline
\hspace*{0.5cm}$\Delta$\hspace*{0.75cm}  & \hspace*{0.55cm}  & $n=1$  & $n=2$  & $n=3$\tabularnewline
\hline\hline
0& $\alpha_n$  & 0.00 (\emph{0})  & 0.25 (\emph{0.25})& 0.44 (\emph{0.444...})  & \tabularnewline
0 & $\beta_n/\phi^2$  & 1.99 (\emph{2})  & 0.99 (\emph{1}) & 0.66 (\emph{0.666...}) & \tabularnewline
\hline
0.5 &  $\alpha_n$ & 0.00 (\emph{0})  & 0.25 (\emph{0.25})  & 0.44 (\emph{0.444...})  & \tabularnewline
0.5 & $\beta_n/\phi^2$  & 1.48 (\emph{1.5}) & 0.75 (\emph{0.75})  & 0.48 (\emph{0.5}) & \tabularnewline
\hline
$\cos(\pi/8)$ & $\alpha_n$  & 0.00 (\emph{0}) & 0.25 (\emph{0.25}) & 0.44 (\emph{0.444...}) &\tabularnewline
$\cos(\pi/8)$ & $\beta_n/\phi^2$  & 1.13 (\emph{1.1428}) & 0.57 (\emph{0.5714})  & 0.35 (\emph{0.3805}) & \tabularnewline
\hline
\end{tabular}

\caption{The values of  $\alpha_n$  and $\beta_n/\phi^2$ obtained by fitting  the
  DMRG data of Fig. 1(a)  to   Eq. (\ref{n3})  for the  spin-$\frac{1}{2}$ XXZ chain
  for  $\Delta=0, 0.5, \cos(\frac{\pi}{8})$.  
The values in  parentheses are the predicted ones in Eq. (\ref{n3}).}
\label{tabI}
\end{table}

{\em Twist fields}. 
Although we have derived the analytic results using the modular properties of non-specialized
characters, we think that  the twist field method of references \cite{CC04,CCD08} 
can be extended   to this case.  This  is suggested by Eq. (\ref{18}), whose r.h.s. is proportional  to 
$L_c(x)^{- \frac{c}{6} (n  - \frac{1}{n} )} \times L_c(x)^{ - \frac{2 K}{n} \phi^2}$.
The first factor comes from the correlator 
$\langle {\cal T}_{n} {\cal T}_{-n} \rangle$ of the twist field ${\cal T}_{\pm n}$ with scaling dimensions 
$\Delta_{{\cal T}_n} = \bar{\Delta}_{{\cal T}_n} = 
\frac{1}{24} ( n - \frac{1}{n})$, and the second factor  corresponds to  the  correlator
$\langle {\cal O}_{\phi,n} {\cal O}_{-\phi,n} \rangle$
of a field ${\cal O}_{\phi,n}$ with scaling dimensions $\Delta_{{\cal O}_{\phi,n}} = \bar{\Delta}_{{\cal O}_{\phi,n}} =  \frac{K}{ 2 n} \phi^2$. 
The field  ${\cal O}_{\phi,n}$ is a generalized string-order parameter with angle $2 \pi \phi$ \cite{DR89},
which for $n=1$ and $\phi= \frac{1}{2}$ has the two point correlator described above \cite{AH92}. 
We expect that the  generalized string-order fields provide an  extension 
of the twist fields, that reminds the ones  used in non unitary
CFTs where the ground state is not the  CFT vacuum \cite{nu1}. 
Double log corrections to the EE have been discussed  in the context
of non unitary CFTs  \cite{nu1,nu2}, and  in 
the non compact Liouville theory with $c=1$  \cite{nu2}.

{\em Conclusions.} We have shown that for  critical Hamiltonians, with a $U(1)$  KM symmetry, 
the bipartite entanglement of the 
projected states exhibits universal properties related to  the underlying CFT
such as the Luttinger parameter $K$, or the level $k$ of the KM algebra  $SU(2)_{k}$. 
The numerical determination of the parameter $K$ using entanglement measures
are quite difficult and  imprecise.  We have presented here a simple way to compute $K$ together with 
the central charge $c$, through the projected density matrices.
We have also derived the probabilities of measuring a given magnetization in a part
of the system,  a problem that is  related to the full counting statistics which 
we generalize to deal with  entanglement effects. 

We believe that the  results obtained  in this Letter can  be measured in
experiments with ultracold atoms.  
For that it is necessary to measure 
${\rm tr}_A  (  e^{  2 \pi i  \phi S^z_A} \rho^n_A  )=\sum_{m}e^{2\pi i\phi m}{\rm tr}_{A}(\rho_{A,m}^{n})$. 
In principle, this quantity can be measured using two different schemes,
proposed  recently in Refs. \onlinecite{measure-opt1} and \onlinecite{measure-opt2}. 
In the scheme of Ref. \cite{measure-opt1}, it is necessary to
build $n$ copies of the state $\rho$. Since ${\rm tr}_{A}(\rho_{A,m}^{n})={\rm tr}_A ( V_n\rho^{\otimes n}_{A,m})$, where
$V_n$  is the  shift operator \cite{measure-opt1}, we only need to measure
the expectation value $<V_n>$ on n copies, for a fixed value of  $m$. Note that
expectation values can be measured in optical lattices \cite{measure-opt1}. 
On the other hand, the scheme proposed in Ref. \onlinecite{measure-opt2} uses a random
measurement protocol in a single copy and for the re-construction it explores
the decomposition of the density matrix  into disjoint blocks with different quantum numbers. This scheme seems
to be a natural route to measure ${\rm tr}_{A}(\rho_{A,m}^{n})$ for a fix
value of $m$. 

Note  that the generalization of our approach  to systems with  higher rank KM algebras like $SU(n)_k$ 
is straightforward and will be reported elsewhere \cite{future}. 
Finally, we would also to point out that  the results obtained in this article
apply only to  critical theories. They can be extended to the massive theories, obtained
by adding relevant perturbation to the critical ones.
The reduced density matrix for an interval whose
size is smaller that the correlation length $\xi$
coincides with the critical one, except that the
cord length $L_c(x)$ is now replaced by the ratio $\xi/a$,
where $a$ is the lattice spacing. The equipartition
of the entanglement entropy, will also  holds
for this more general class of models. 
\vspace{0.2cm}

{\em Noted added.} After completion of this work we were informed on the references
\cite{LR13,GS17} that also consider  the problem studied here.


{\it Acknowledgements.} 
\begin{acknowledgments}
GS would like to thanks A. Ludwig for a useful discussion. 
We also acknowledge conversations with 
L. Balents, E. Fradkin, J.I. Latorre, E. L\'opez,  
 J. Rodr\'{\i}guez-Laguna,   
M. Srednicki, H. Tu, W. Witczak-Krempa and G. Vidal. 
We acknowledge financial support from the Brazilian agencies FAPEMIG, FAPESP, and CNPq, 
the grants FIS2015-69167-C2-1-P, QUITEMAD+ S2013/ICE-2801 and SEV-2016-0597 of the "Centro de Excelencia
Severo Ochoa" Programme.  This research was also 
supported in part  by  the  Grant No. NSF PHY17-48958.
\end{acknowledgments}

\end{document}